\documentstyle[epsfig,sprocl]{article}
\begin{document}
\thispagestyle{empty}
\begin{flushright}
\large
RIKEN-AF-NP-293 \\ July 1998
\end{flushright}
\vspace{0.65cm}

\begin{center}
\LARGE
{\bf RHIC SPIN PHYSICS}

\vspace{0.1cm}

\vspace{1.2cm}
\Large
N.~HAYASHI, Y.~GOTO and N.~SAITO \\

\vspace{0.5cm}
\large
Radiation Laboratory, \\
\vspace{0.1cm}
The Institute of Physical and Chemical Research (RIKEN),\\
\vspace{0.1cm}
2-1 Hirosawa, Wako, 351-0098 JAPAN\\
\vspace{0.1cm}
E-mail: nhayashi@riken.go.jp \\
\vspace{2.4cm}
{\bf Abstract} 
\end{center}
\vspace*{0.5cm}

\noindent
We describe the foreseen spin physics of the polarized
proton-proton collider at RHIC. 
The study of the spin structure of the nucleon at RHIC is 
unique and complementary to the traditional 
polarized DIS experiments. 
The sensitivities of the gluon polarization measurement
via prompt photon production and anti-quark polarization 
measurement using $W$ production with the PHENIX detector
system are reviewed. 
\normalsize

\vspace{3.0cm}
\noindent
{\it Talk presented at the 6th International Workshop on
Deep Inelastic Scattering and QCD (DIS98), Brussels, Belgium,
April 4-8, 1998.}
\vfill

\setcounter{page}{0}
\newpage
\font\eightrm=cmr8

\bibliographystyle{unsrt} 

\arraycolsep1.5pt

\def\Journal#1#2#3#4{{#1} {\bf #2}, #3 (#4)}

\def\NCA{\em Nuovo Cimento}
\def\NIM{\em Nucl. Instrum. Methods}
\def\NIMA{{\em Nucl. Instrum. Methods} A}
\def\NPB{{\em Nucl. Phys.} B}
\def\PLB{{\em Phys. Lett.}  B}
\def\PRL{\em Phys. Rev. Lett.}
\def\PRD{{\em Phys. Rev.} D}
\def\ZPC{{\em Z. Phys.} C}

\def\st{\scriptstyle}
\def\sst{\scriptscriptstyle}
\def\mco{\multicolumn}
\def\epp{\epsilon^{\prime}}
\def\vep{\varepsilon}
\def\ra{\rightarrow}
\def\ppg{\pi^+\pi^-\gamma}
\def\vp{{\bf p}}
\def\ko{K^0}
\def\kb{\bar{K^0}}
\def\al{\alpha}
\def\ab{\bar{\alpha}}
\def\be{\begin{equation}}
\def\ee{\end{equation}}
\def\bea{\begin{eqnarray}}
\def\eea{\end{eqnarray}}
\def\CPbar{\hbox{{\rm CP}\hskip-1.80em{/}}}



\title{RHIC SPIN PHYSICS}

\author{N.~HAYASHI, Y.~GOTO and N.~SAITO}
\address{Radiation Laboratory, \\
The Institute of Physical and Chemical Research (RIKEN),\\
2-1 Hirosawa, Wako, 351-0098 JAPAN\\
E-mail: nhayashi@riken.go.jp}


\maketitle\abstracts{ 
We describe the foreseen spin physics of the polarized
proton-proton collider at RHIC. 
The study of the spin structure of the nucleon at RHIC is 
unique and complementary to the traditional 
polarized DIS experiments. 
The sensitivities of the gluon polarization measurement
via prompt photon production and anti-quark polarization 
measurement using $W$ production with the PHENIX detector
system are reviewed. 
}

\section{Introduction}

The spin-dependent structure functions of the nucleon have been
studied extensively in polarized deep inelastic scattering (pDIS)
experiments in the last two decades. These experiments provide a lot
of information particularly the quark spin contributions to the
nucleon spin. 
However, the size of the gluon spin contribution is still not
well constrained.
Moreover, pDIS inclusive reaction alone is not possible to separate
individual contributions of each quark flavor or that of the
anti-quark.

RHIC/Spin experiments are the new type of the experiments which
use polarized proton-proton collisions at very high energy.
After a brief overview of the accelerator and the detectors aspects,
we discuss measurements of the gluon polarization via the prompt
photon production and how to
disentangle the contribution from different (anti-) quark flavor via
$W$ production.
Although this facility is to provide possibilities of various studies,
we will concentrate on the topics mentioned above in this paper.

\section{RHIC and Detectors}

The Relativistic Heavy Ion Collider (RHIC) is now under construction
at Brookhaven National Laboratory (BNL). The original aim of the RHIC
is to explore the quark-gluon plasma with the heavy ion collisions.
In addition the proposal of the spin physics program at
RHIC~\cite{RSCR5} has been approved.
Special sets of a helical dipole magnets, called the Siberian snake
and the spin rotators, will be installed into both of the RHIC rings in
order to preserve the polarization of the proton and manipulate the
spin directions at two colliding points.
Necessary improvements at the polarized ion source or the polarized
beam acceleration at the AGS (RHIC injector) 
are also underway. 
The center of mass energy ($\sqrt{s}$) of the RHIC is in the range of
$50$ 
to $500$ GeV with the high beam polarization (about 70\%). A
high luminosity is also expected, $8\times 10^{31}$ ($2\times
10^{32}$) cm$^{-2}$ s$^{-1}$ at $\sqrt{s}=200$ ($500$) GeV. 
The integrated luminosities aiming to accumulate are 320 and 800
pb$^{-1}$ at $\sqrt{s}=200$ and $500$ GeV, respectively.
Sensitivity studies have been done by assuming these luminosities. 

PHENIX and STAR are two major international collaborations of over 400
 physicists each.
%
The PHENIX detector consists of Central Arm and two Muon
Arms. Electro-magnetic (EM) calorimeter is in Central Arm and covers
$|\eta|<0.35$ in rapidity and half azimuthal range ($90^{\circ} \times
2$). Its acceptance is limited but it has fine granularity
($\Delta\eta \simeq \Delta\phi \simeq 0.01$). Tracking elements
cover the same acceptance range. Muon Arms cover full azimuthal angle
and forward rapidity regions ($1.2<|\eta|<2.4$).
The heart of STAR detector is Time-Projection-Chamber (TPC). It covers
wide rapidity region ($-2<\eta<2$) together with Silicon Vertex
Detector (SVD). 
The EM calorimeter consists of the barrel one ($-1<\eta<1$) and one
endcap ($1<\eta<2$). 
Jet can be identified with its larger acceptance.

\section{Gluon Polarization Measurement}

The prompt photon production in $pp$ collisions is one of the best
methods to study the gluon. 
Two processes contribute to the photon
production in the leading order (LO), namely the gluon Compton
scattering ($g+q\rightarrow \gamma+q$) and the annihilation process
($q+\bar{q}\rightarrow\gamma+g$). At RHIC energy 
the fraction of the Compton scattering is estimated to be $80\sim 90 \%$.
Neglecting the annihilation contribution,
the double-longitudinal asymmetry of the prompt photon production is expressed in eq.(\ref{phtn-asym}), 
\begin{equation}
A_{LL}={\Delta G(x_1)\over G(x_1)}{g_1^p(x_2)\over F_1^p(x_2)}\hat{a}_{LL}(g+q\rightarrow\gamma+q);
{g_1(x)\over F_1(x)}\simeq {\sum_i e_i^2\Delta q_i(x)\over \sum_i e_i^2 q_i(x)}
\label{phtn-asym}
\end{equation}
where $g_1^p(x)\over F_1^p(x)$ is the measured proton asymmetry in pDIS
and ${\Delta G(x)\over G(x)}$ is the gluon polarization.
The subprocess asymmetry $\hat{a}_{LL}$ is calculated in QCD.

\begin{figure}[t]
\begin{center}
\epsfig{file=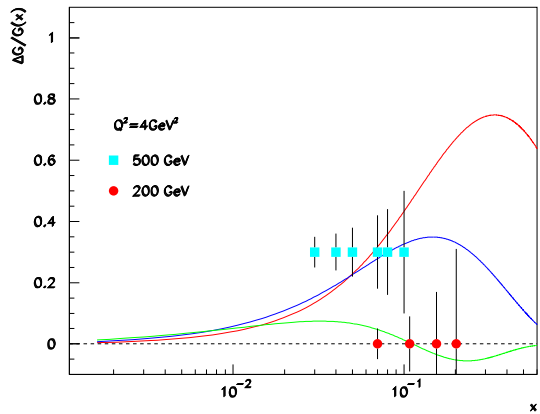,width=8cm}
\caption[]{Expected gluon polarization measurement via the prompt
photon measurement with PHENIX detector. Three solid lines indicate
the three theoretical models of T.Gehrmann and W.J.Stirling~\cite{GS96}.}
\label{dg-over-g}
\end{center}
\end{figure}

In general, extracting the prompt photon signal from the large
background, mainly from $\pi^0 \rightarrow 2 \gamma$, is not simple.
Moreover, reconstruction of parton
level kinematics, namely the determination of $x$ carried by gluon, is
challenging.
Studies to resolve those problems are
in progress in both PHENIX and STAR.

The finely segmented PHENIX EM calorimeter is advantageous in
identifying
$\pi^0$. Its limited solid angle, however, allows only inclusive
measurement,
which selects parton level kinematics rather vaguely. Our simulation
study shows
strong correlation between measured $p_T$ and $x$ carried by the gluon.
The average value of $x$ in each $p_T$ bin is smaller than
$x_T = 2\cdot p_T/\sqrt{s}$,
due to (i) the presence of intrinsic $k_T$ and (ii) imbalance of initial
momenta carried by gluon and quark (quark tends to carry larger momentum
than gluon in this
$x$ region) .
Sensitivities on the gluon polarization, $\frac{\Delta G(x)}{G(x)}$ have
been estimated using eq. (1) with the expected
yield, measured $\frac{g_1^p(x)}{F_1^{p}(x)}$, and theoretically
calculated $\hat{a}_{LL}$. Results are shown Figure 1. Open and closed circles
show probed $x$ and error bars show projected statistical errors.
The errors are estimated to be 5-30\% at $\sqrt{s}=200$~GeV, and
5-20\% at 500~GeV in the $x-$bins. Further systematic studies are
underway.

On the other hand, STAR has a large photon acceptance and jet detection
capability. Thus more strict reconstruction of the parton level
kinematics
is possible, although uncertainty in assigning resolved $x_1$ and $x_2$
 to
gluon and quark still remains. More detailed description can be found in
ref.\cite{Heppelmann}.

\section{Quark and Anti-quark Polarization Measurement}



The charged weak boson production at the polarized RHIC 
provides unique information on the spin structure studies.
Because of its parity violating nature, identifying $W^+$
or $W^-$ defines the helicity of the parents quark and
anti-quark. At the same time, it gives a strong constraint on the
quark and anti-quark flavors.
For example, the single spin asymmetry of $W^+$ production is
expressed as
\begin{equation}
A_{L}^{W^+}=
{\Delta u(x_1) \bar{d}(x_2) - \Delta\bar{d}(x_1) u(x_2) \over
u(x_1) \bar{d}(x_2) + \bar{d}(x_1) u(x_2) },
\label{alwp}
\end{equation}
where $u$ ($\bar{d}$) is the unpolarized $u$- ($\bar{d}$-) quark 
distribution in the proton and $\Delta u$ ($\Delta\bar{d}$) is the
polarized $u$- ($\bar{d}$-) quark distribution. $x_1$ ($x_2$) is
the fraction of the (un)polarized proton carried by the parton.
%
In the high rapidity limit of $W^+$, $A_{L}^{W^+}$ becomes approximately
$\Delta u(x)\over u(x)$ and in the other side of the limit, $\Delta
\bar{d}(x) \over \bar{d}(x)$ becomes dominant.

The $W^{\pm}$ can be detected via the leptonic decay, either $\mu^{\pm}$
(PHENIX Muon Arm) or $e^{\pm}$ (PHENIX/STAR EM calorimeter).
In the case of $\mu$, it has been checked that high $p_T$ $\mu$ above
20 GeV/$c$ is dominated by $W\rightarrow \mu+\nu$ decay.
At PHENIX Muon Arm we expect about 5000 events for each $W^+$ and
$W^-$ with this cut, which corresponds to 2\% of statistical error for
$A_L$.
It should be noted that geometrical acceptance for $W^+$ is quite
different from  $W^-$ due to the decay angle distribution.

Since the decay neutrino cannot be detected, it is hard to
determine parton kinematics exactly. However, (in $p_T(W)\rightarrow
0$ limit) the energy of the decay muon and the parent quark's $x$
are highly correlated and it gives a good estimate of $x$.
\begin{figure}[t]
\begin{center}
\epsfig{file=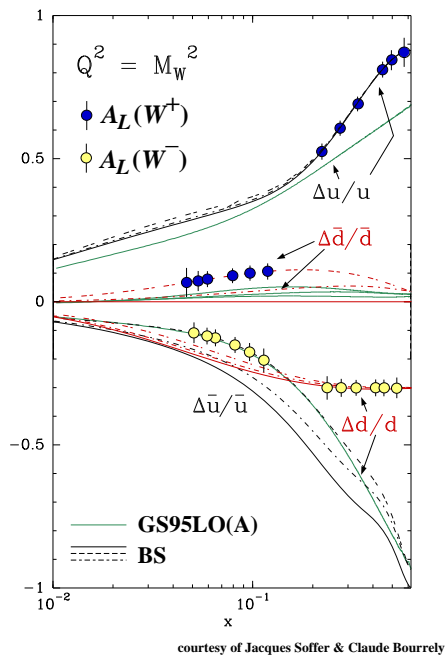,width=6cm}
\caption[]{$\Delta q$, $\Delta\bar{q}$ sensitivity at RHIC.
Various theoretical models%
~\cite{GS96,BS}
are also plotted on the figure.}
\label{dqqbar-w}
\end{center}
\end{figure}
The sensitivities of the various quark polarization via $W$ production
is summarized in Fig.\ref{dqqbar-w}. 
Error bars indicate the expected statistical errors. As shown in this
figure, our measurement is very sensitive to the quark spin.
Further systematics
studies have to be done in order to minimize uncertainties.

\section{Summary}

The measurement of the spin structure function of the proton at RHIC
were discussed. 
Two major subjects were addressed:
$\Delta G(x)$ measurements via the prompt photon and
$\Delta\bar{q}(x)$ measurements through $W$ production.
The statistical error of ${\Delta G(x)\over G(x)}$ with PHENIX is
5-30\% and that of the $W$ production is 2\%.


According to the current schedule, RHIC will be turned on in the
beginning of 1999 and the first heavy ion physics program is foreseen
in the fall of 1999. A commissioning of the polarized proton beam is
planed in the running period which starts 1999 fall.
The first polarized proton physics will start in the running period which
begins year 2000 after the commissioning.

\section*{Acknowledgments}

The authors would like to thank to both PHENIX and STAR
collaborations, particularly, Drs .G.~Bunce, M.~Tannenbaum and
A.~Ogawa for the discussions related to this work.

\end{document}